\documentclass[a4paper,11pt]{article}
\usepackage[german,american]{babel}
\usepackage[a4paper,top=3.0cm,bottom=2.5cm,left=2.5cm,right=2.5cm,marginparwidth=1.75cm]{geometry}
\usepackage[numbers]{natbib}
\usepackage{graphicx}
\usepackage{amsmath}
\usepackage{amssymb}
\usepackage{booktabs}
\usepackage{array}
\usepackage{multirow}
\usepackage{xcolor}
\usepackage{siunitx}
\usepackage{chemformula}
\usepackage{url}
\usepackage{hyperref}
\usepackage{float}
\usepackage{tabularx}
\usepackage{seqsplit}
\title{Escaping the Hydrolysis Trap: An Agentic Workflow for Inverse Design of Durable Photocatalytic Covalent Organic Frameworks}
\author{
    Iman Peivaste$^{1,2,*}$,
    Nicolas D. Boscher$^{1}$,
    Ahmed Makradi $^{1}$,
    Salim Belouettar $^{1}$ \\
\\
\footnotesize $^1$ Luxembourg Institute of Science and Technology (LIST), 5,\\
\footnotesize Avenue des Hauts-Fourneaux, Esch-sur-Alzette, 4362, Luxembourg\\
\\
\footnotesize $^2$ Department of Physics and Materials Science,\\
\footnotesize University of Luxembourg, L-4365 Esch-sur-Alzette, Luxembourg \\
\\
\footnotesize $^*$Corresponding Author: iman.peivaste@list.lu
}
\date{\today}

\begin{document}
\maketitle

\begin{abstract}
Covalent organic frameworks (COFs) are promising photocatalysts for solar hydrogen production, yet the most electronically favorable linkages, imines,  hydrolyze rapidly in water, creating a stability–activity trade-off that limits practical deployment. Navigating the combinatorial design space of nodes, linkers, linkages, and functional groups to identify candidates that are simultaneously active and durable remains a formidable challenge. Here we introduce \textbf{Ara}, a large-language-model (LLM) agent that leverages pretrained chemical knowledge, donor–acceptor theory, conjugation effects, and linkage stability hierarchies, to guide the search for photocatalytic COFs satisfying joint band-gap, band-edge, and hydrolytic-stability criteria. Evaluated against random search and Bayesian optimization (BO) over a space consisting of candidates with various nodes, linkers, linkages, and r-groups, screened with a GFN1-xTB fragment pipeline, Ara achieves a 52.7\% hit rate (11.5$\times$ random, $p = 0.006$), finds its first hit at iteration~12 versus~25 for random search, and significantly outperforms BO ($p = 0.006$). Inspection of the agent's reasoning traces reveal interpretable chemical logic: early convergence on vinylene and $\beta$-ketoenamine linkages for stability, node selection informed by electron-withdrawing character, and systematic R-group optimization to center the band gap at 2.0~eV. Exhaustive evaluation of the full search space uncovers a complementary exploitation-exploration trade-off between the agent and BO, suggesting that hybrid strategies may combine the strengths of both approaches. These results demonstrate that LLM chemical priors can substantially accelerate multi-criteria materials discovery.
\end{abstract}

\section*{Introduction}

Photocatalytic water splitting driven by solar energy offers a sustainable route to hydrogen fuel, and the search for efficient, stable photocatalysts remains one of the central challenges in materials science \cite{chen2017particulate,wang2019particulate, tao2022recent, qi2018solar}. Covalent organic frameworks (COFs), crystalline, porous polymers assembled from organic building blocks through reversible covalent bonds, have emerged as attractive photocatalyst candidates owing to their tunable band gaps, high
surface areas, and modular synthetic chemistry \cite{ding2013covalent,cote2005porous,geng2020covalent}. In principle, the band gap and conduction-band minimum (CBM) of a COF can be rationally engineered by selecting appropriate electron-donating and
electron-withdrawing building blocks, making COFs a natural platform for inverse design \cite{stegbauer2014hydrazone,vyas2015tunable}. In practice, however, a fundamental tension exists between electronic performance and chemical durability: the imine (C=N) linkages that dominate the COF literature are susceptible to hydrolysis under
the aqueous, often acidic conditions of photocatalytic operation \cite{lyu2019porous,xu2015stable}. This "hydrolysis trap" means that many COFs with promising band gaps degrade before they can function.

The design space for COF photocatalysts is vast. Combining a modest library of
seven trigonal nodes, nineteen ditopic linkers, four linkage chemistries, and
ten aromatic R-group substituents yields 820 compatibility-constrained
candidates, and realistic libraries can be orders of magnitude larger
\cite{ongari2019building}. Density functional theory (DFT) calculations, while accurate,
require hours per candidate for periodic structures, making exhaustive screening
prohibitively expensive \cite{haase2020solving,jain2016computational}. Even semiempirical methods such as
the GFN$n$-xTB family \cite{grimme2017robust} require minutes per geometry optimization,
and the multi-criteria nature of the problem, simultaneous satisfaction of
band-gap, band-edge, and stability requirements, rules out simple ranking on
a single descriptor.

Several computational strategies have been proposed to navigate large materials
design spaces. High-throughput virtual screening enumerates and evaluates
candidates sequentially but scales poorly beyond a few hundred evaluations
\cite{hachmann2011harvard}. Bayesian optimization (BO) places a Gaussian process surrogate
over molecular fingerprints and uses an acquisition function to balance
exploration and exploitation \cite{shields2021bayesian,griffiths2020constrained}; however, BO treats molecules
as feature vectors and cannot reason about categorical chemical concepts such
as linkage hydrolysis or band-gap tuning via donor--acceptor engineering. Machine-learning surrogate
models accelerate screening once trained \cite{butler2018machine,moosavi2020role, peivaste2025artificial, peivaste2025hype} but face a
cold-start problem when no labelled data exist for a new materials class or
property combination.

Large language models (LLMs) offer a qualitatively different approach. Trained
on the scientific literature, LLMs encode chemical knowledge, donor--acceptor
theory, and conjugation effects, functional-group reactivity that can serve as a
prior for materials design without task-specific training data \cite{jablonka2024leveraging,zheng2023chatgpt,boiko2023autonomous}.
Recent work has explored LLMs as assistants for retrosynthesis \cite{schwaller2019molecular},
molecular property prediction \cite{ramos2023bayesian}, deriving design rules \cite{peivaste2026chemnavigator}, and reaction optimization \cite{m2024augmenting},
but their application to multi-criteria materials discovery with explicit chemical
reasoning remains largely unexplored. Here we present \textbf{Ara}, an LLM-guided
agent built on Google's Gemini model \cite{team2023gemini} that navigates COF design space
using an iterative reasoning loop: at each step, the agent receives feedback on
its previous selections (band gap, CBM, stability score) and proposes a new
candidate with a brief chemical justification. We pair the agent with a
fragment-based screening pipeline that uses GFN1-xTB to compute vertical
ionization potentials and electron affinities, calibrated against periodic DFT
band gaps for 13 COFs (Spearman $\rho = 0.71$, $p = 0.006$). A composite
stability index ($S_\text{CSI}$) that integrates linkage chemistry, steric
shielding, and hydrophobicity encodes the stability axis of the design problem.

We evaluate Ara against random search and BO over 200 iterations across five independent seeds. The agent achieves a 52.7\% hit rate, 11.5 times that of random search ($p = 0.006$, Mann-Whitney U, one-sided), and significantly outperforms BO ($p = 0.006$) on cumulative hits. Analysis of the agent's reasoning traces reveals interpretable chemical logic: convergence on
non-hydrolyzable vinylene linkages, avoidance of excessively electron-withdrawing nodes, and systematic R-group tuning. Exhaustive
evaluation of all candidates identifies 38 ground-truth hits, enabling an oracle analysis that exposes a complementary exploitation-exploration trade-off: the agent excels at rapidly identifying high-quality candidates within a budget-constrained campaign, while BO maps a larger fraction of the hit landscape through systematic exploration. Together, these results demonstrate that LLM chemical priors can substantially accelerate multi-criteria materials discovery and that agent-based and surrogate-based strategies serve
complementary experimental goals.

\section*{Results}

\begin{figure}[H]
\centering
\includegraphics[width=1\linewidth]{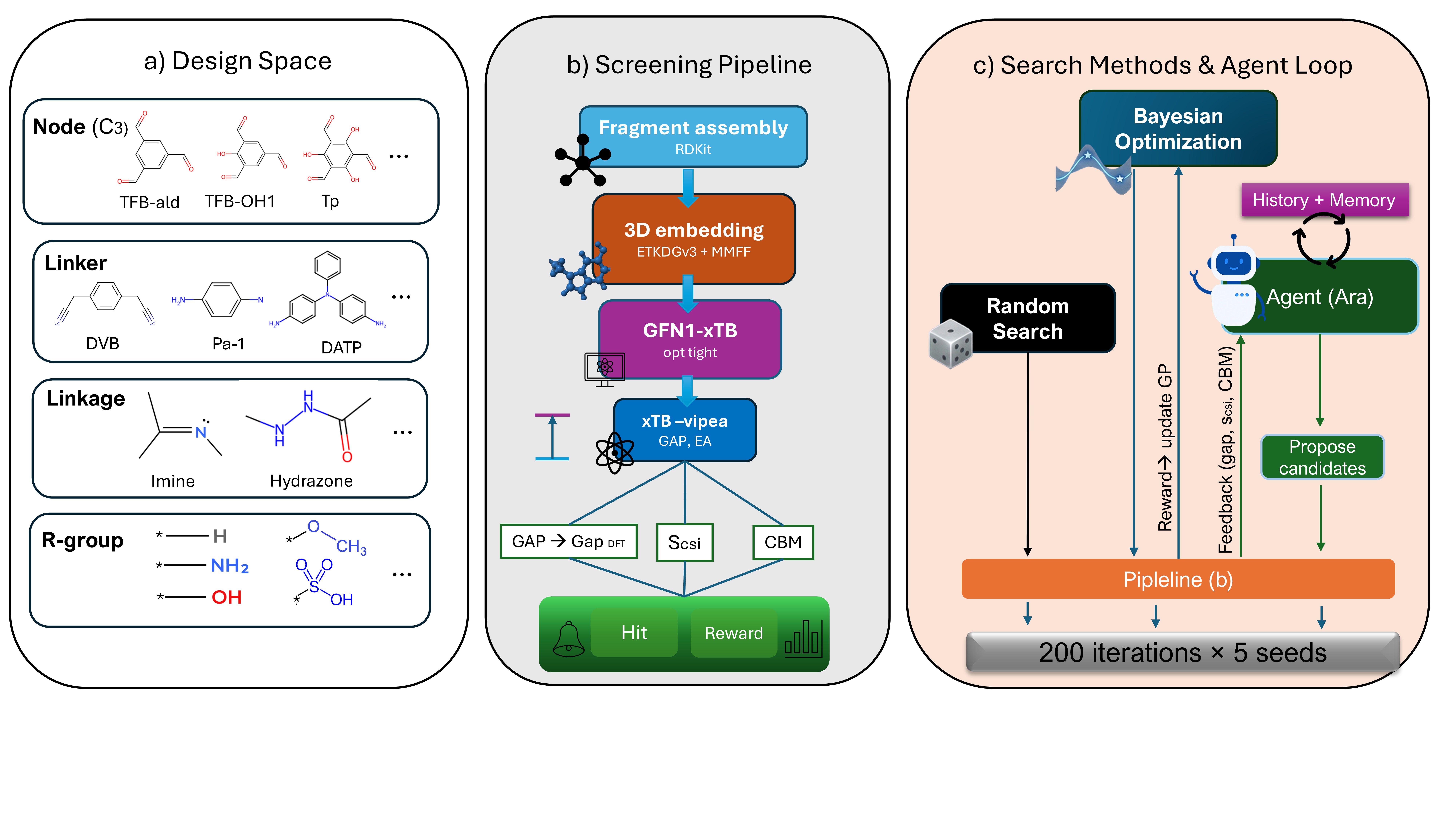}
\caption{Overview of the Ara agentic workflow for COF photocatalyst discovery. (a) The combinatorial design space comprises 820 candidates formed from 7 trigonal nodes, 19 ditopic linkers, 4 linkage chemistries of varying hydrolytic stability, and 10 aromatic R-group substituents, subject to chemical compatibility constraints. (b) Each candidate is evaluated through a fragment-based screening pipeline: a node–linker–node repeat unit is assembled via RDKit, embedded in 3D, optimized with GFN1-xTB, and scored for band gap (IP$-$EA, mapped to the DFT scale via a calibrated transfer function), conduction-band minimum (CBM), and composite stability index (S\_CSI). A candidate is classified as a hit if it satisfies all three criteria simultaneously. (c) Three search strategies are compared: random sampling, Bayesian optimization with a Gaussian process surrogate on Morgan fingerprints, and the Ara LLM agent, which iteratively proposes candidates with explicit chemical reasoning, receives quantitative feedback, and refines its selections over 200 iterations.}
\label{fig:pipeline}
\end{figure}

\subsection*{Calibration validates fragment-based screening}

\begin{figure}[H]
    \centering
    \includegraphics[width=0.75\linewidth]{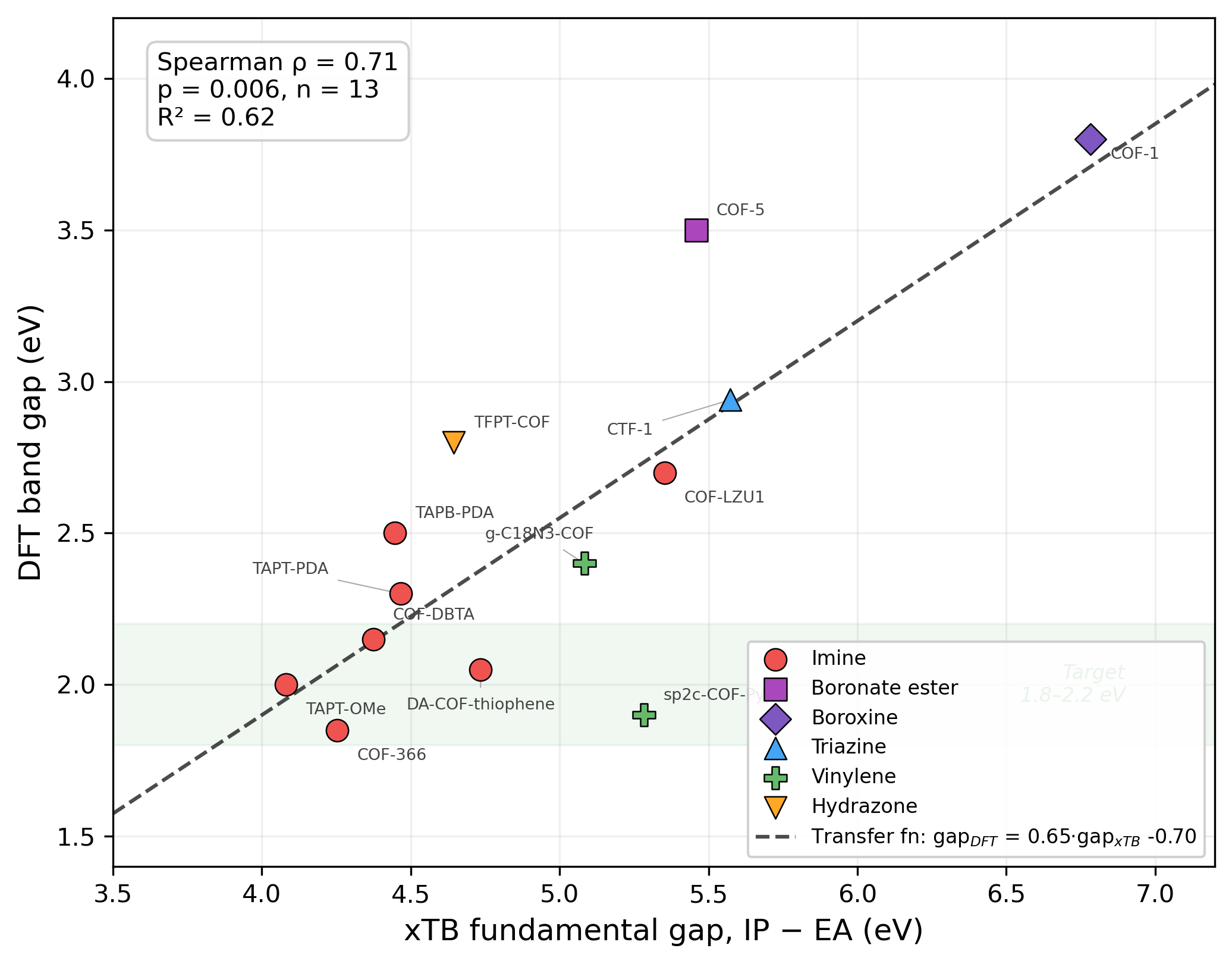}
    \caption{Scatter plot of xTB (IP$-$EA) fundamental gap versus DFT band gap for 13 COFs spanning six linkage types, with the linear transfer function overlaid. The calibration set includes boronate ester, boroxine, and triazine linkage types not present in the search space to broaden the range of the transfer function. Note that CTF-1 employs triazine ring formation as the linkage chemistry, distinct from triazine-containing nodes (e.g., TFPT-ald, TAPT) that connect via imine or other bond types.}
    \label{fig:calib}
\end{figure}

The fragment-based fundamental gap (IP$-$EA) from GFN1-xTB correlated significantly with periodic DFT band gaps across six linkage types (Spearman $\rho = 0.71$, $p = 0.006$, $n = 13$; Supplementary Table~S2), validating the screening pipeline for comparative ranking of candidates (Fig.~\ref{fig:calib}). Three methodological pivots were required to reach this result. First, initial attempts using $2 \times 2$ periodic clusters inverted the gap ranking of known COFs, motivating a switch to repeat-unit fragments (${\sim}50$--80 atoms) that preserved the correct ordering \cite{onida2002electronic}. Second, HOMO--LUMO gaps from xTB Single-point calculations showed no correlation with DFT values ($\rho = -0.18$, $p = 0.70$ for a seven-COF imine-only subset), whereas fundamental gaps computed as IP$-$EA via the delta-SCF protocol (\texttt{xtb -{}-vipea}) \cite{grimme2017robust} recovered the expected trend, a finding consistent with the known inadequacy of Kohn-Sham eigenvalue differences for organic framework quasiparticle gaps. Third, five $\beta$-ketoenamine COFs were excluded from calibration because their fragment SMILES encoded the imine tautomeric form rather than the correct keto-enamine structure, producing erroneous electronic structures. A linear transfer function,
\begin{equation}
  \text{gap}_\text{DFT} \approx 0.65 \times \text{gap}_\text{xTB} - 0.70
  \quad (R^2 = 0.62),
  \label{eq:transfer}
\end{equation}
mapped xTB fundamental gaps onto the DFT scale for downstream screening.

These results establish that fragment-level IP$-$EA calculations from GFN1-xTB provide a computationally inexpensive yet reliable proxy for periodic DFT band gaps. The transfer function and CBM calibration (see Methods) formed the quantitative backbone of all subsequent screening.

\subsection*{Agent achieves superior sample efficiency}

We compared three search methods, random search, Bayesian optimization (BO), and Ara (LLM agent), over 200 iterations with five independent random seeds each. A candidate was classified as a ``hit'' if it simultaneously satisfied
three criteria: a corrected band gap between 1.8 and 2.2~eV, CBM below 0~V, and composite stability index $S_\text{CSI} \geq 0.7$ (see Methods). Table~\ref{tab:comparison} summarizes the results.

\begin{table}[H]
\centering
\caption{Comparative study results (mean $\pm$ s.d.\ over five seeds).}
\label{tab:comparison}
\small
\begin{tabular}{lrrrrr}
\toprule
\textbf{Method} & \textbf{Cum.\ hits} & \textbf{Hit rate (\%)} &
\textbf{First hit (iter.)} & \textbf{Best reward} & \textbf{Success rate (\%)} \\
\midrule
Random & $9.2 \pm 1.9$  & 4.6  & $25 \pm 16$ & $0.903 \pm 0.012$ & 82.1  \\
BO     & $28.2 \pm 3.8$ & 14.1 & $22 \pm 31$ & $0.921 \pm 0.000$ & 100.0 \\
Agent  & $105.4 \pm 40.9$ & 52.7 & $12 \pm 13$ & $0.895 \pm 0.016$ & 95.7  \\
\bottomrule
\end{tabular}
\end{table}

\begin{figure}[H]
\centering
    \includegraphics[width=0.75\linewidth]{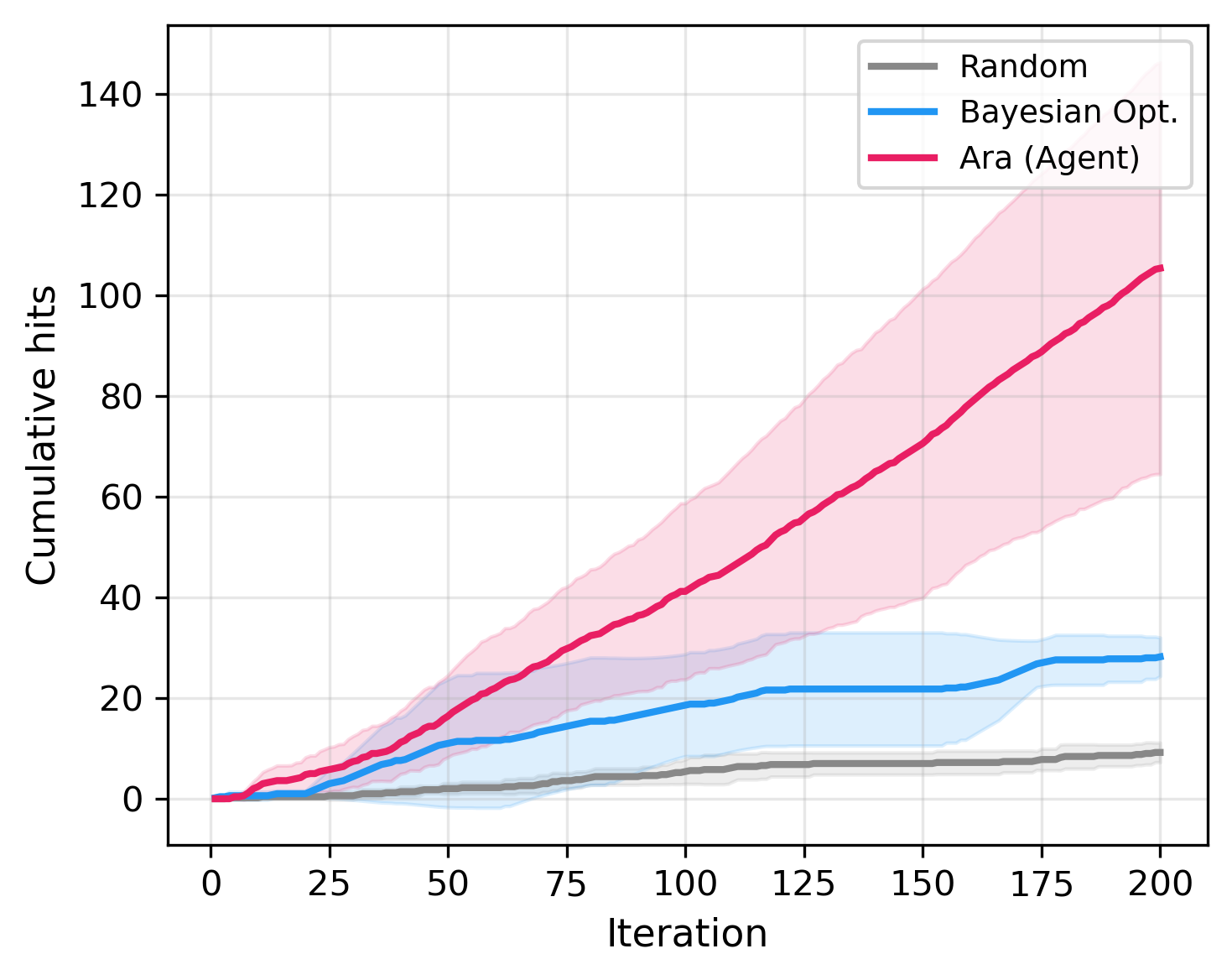}
\caption{Cumulative hits versus iteration for random search (grey), Bayesian optimization (blue), and Ara (agent, pink). Shaded regions indicate mean~$\pm$~s.d.\ across five random seeds (200 iterations per run). A hit is defined as a candidate satisfying 1.8--2.2\,eV band gap, CBM~$< 0$~V, and S\textsubscript{CSI}~$\geq 0.7$. The agent outperforms both random search and BO in cumulative hits and discovers its first hit sooner.}
\label{fig:curves}
\end{figure}

The Ara agent discovered hits at 11.5 times the rate of random search ($p = 0.006$, Mann--Whitney U, one-sided), with the earliest median first-hit iteration among all methods (iteration~12 versus~25 for random and~22 for BO; Table~\ref{tab:comparison},
Fig.~\ref{fig:curves}). The cumulative hit curves (Fig.~\ref{fig:curves}) show that the agent's advantage is not a transient early effect: the agent curve rises steeply from approximately iteration~15 and continues to dominate both baselines through iteration~200. The agent also significantly outperformed BO on cumulative hits ($U = 25$, $p = 0.006$). BO, however, achieved the highest single-candidate reward (0.921), reflecting its strength at scalar optimization through Gaussian process extrapolation, whereas the agent's advantage lay in the sustained identification of candidates meeting all three hit criteria simultaneously.

Despite 23\% of agent iterations falling back to random selection due to model-output parsing failures, the agent achieved an 11.5-fold improvement over pure random search. This implies that the ${\approx}77\%$ of successfully guided iterations were dramatically more effective than chance, and that even partial LLM guidance provides substantial benefit. The xTB evaluation success rate was 95.7\% for agent-selected candidates versus 82.1\% for random, suggesting that the agent's chemistry-aware selections also avoided combinations prone to assembly
or convergence failure.

The agent's decisive advantage in cumulative hits, combined with earlier first-hit discovery, establishes that LLM-guided search substantially outperforms both random and surrogate-based strategies for this multi-criteria design task.

\subsection*{Chemical reasoning drives agent performance}

To understand why the agent outperforms BO, we collected detailed reasoning traces from a dedicated 20-iteration run with increased output-token limits (see Methods). The traces reveal a coherent chemical strategy that unfolds over the first dozen iterations (Table~\ref{tab:traces}).

\begin{table*}[t]
\centering
\caption{Selected reasoning traces from the agent (20-iteration dedicated run,
seed~42). ``Hit'' column marks candidates meeting all three criteria.}
\label{tab:traces}
\footnotesize
\setlength{\tabcolsep}{5pt}
\begin{tabularx}{\textwidth}{cp{4.6cm}rrcX}
\toprule
\textbf{Iter.} & \textbf{Selection} &
\textbf{Gap (eV)} & \textbf{$S_\text{CSI}$} & \textbf{Hit} &
\textbf{Reasoning (abbreviated)} \\
\midrule
0  & Tp $+$ BD $+$ $\beta$-keto.\ $+$ H
   & 2.61 & 0.67 & --- &
   $\beta$-Ketoenamine for stability via irreversible tautomerization. \\
1  & Tp $+$ BTDA $+$ $\beta$-keto.\ $+$ H
   & 2.50 & 0.64 & --- &
   Benzothiadiazole acceptor to narrow the gap into the target range. \\
2  & TFPT-ald $+$ DVB $+$ vinyl.\ $+$ H
   & 1.53 & 0.73 & --- &
   Triazine acceptor $+$ vinylene for non-hydrolysable stability; gap too low. \\
3  & TFB $+$ DVB $+$ vinyl.\ $+$ H
   & 2.13 & 0.73 & \textbf{Hit} &
   Electron-neutral TFB replaces strong acceptor TFPT-ald to widen the gap. \\
4  & TFP-OH1 $+$ DVB $+$ vinyl.\ $+$ H
   & 2.11 & 0.73 & \textbf{Hit} &
   Mild e$^{-}$-donor OH narrows the gap slightly from 2.13 toward 2.0~eV. \\
5  & TFP-OH1 $+$ DVB $+$ vinyl.\ $+$ OMe
   & 2.02 & 0.75 & \textbf{Hit} &
   Methoxy donor narrows the gap toward the 2.0~eV ideal. \\
8  & TFB $+$ DVB $+$ vinyl.\ $+$ OMe
   & 2.03 & 0.73 & \textbf{Hit} &
   Same strategy applied to TFB scaffold; gap centres at 2.0~eV. \\
11 & TFB $+$ DVB $+$ vinyl.\ $+$ Me
   & 2.12 & 0.73 & \textbf{Hit} &
   Weak donor; gap remains in target window. \\
\bottomrule
\end{tabularx}
\end{table*}

The reasoning traces reveal that the agent's selections were guided by explicit chemical logic, band-gap tuning via donor--acceptor engineering, linkage stability hierarchies, and iterative refinement based on quantitative feedback, rather than pattern matching or memorization of training data. The agent began by testing $\beta$-ketoenamine linkages (iterations 0--1), explicitly citing their hydrolytic stability via irreversible tautomerization. Although these candidates had adequate stability indices, their band gaps exceeded the 2.2~eV upper bound. At iteration~2, the agent switched to a vinylene linkage with the strongly electron-withdrawing TFPT-ald node, reasoning that the donor--acceptor combination would narrow the gap while providing non-hydrolysable C$=$C bonds. When the resulting gap (1.53~eV) fell below the 1.8~eV threshold, the agent corrected course at iteration~3 by replacing TFPT-ald with the electron-neutral TFB node, a move it justified as reducing excessive acceptor character, widening the gap to 2.13~eV and producing the first hit.

From iteration~4 onward, the agent pursued a systematic optimization strategy within the vinylene-linked scaffold family. It first tested TFP-OH1, a node with a mild electron-donating hydroxyl group, observing that the gap narrowed slightly
(2.11~eV). It then explored R-group substitutions, OMe (gap 2.02~eV), NH$_2$ (2.10~eV), OH (2.14~eV), Me (2.12~eV), using donor and acceptor electronic effects to fine-tune the gap toward the 2.0~eV optimum. Each selection was accompanied by a reasoning statement that referenced the previous result and predicted the direction of the gap change, and these predictions were frequently correct. This iterative refinement loop, propose, evaluate, reason, adjust, is a qualitatively different mode of search from either random
sampling or fingerprint-based acquisition functions.

\begin{figure}[H]
\centering
    \includegraphics[width=0.9\linewidth]{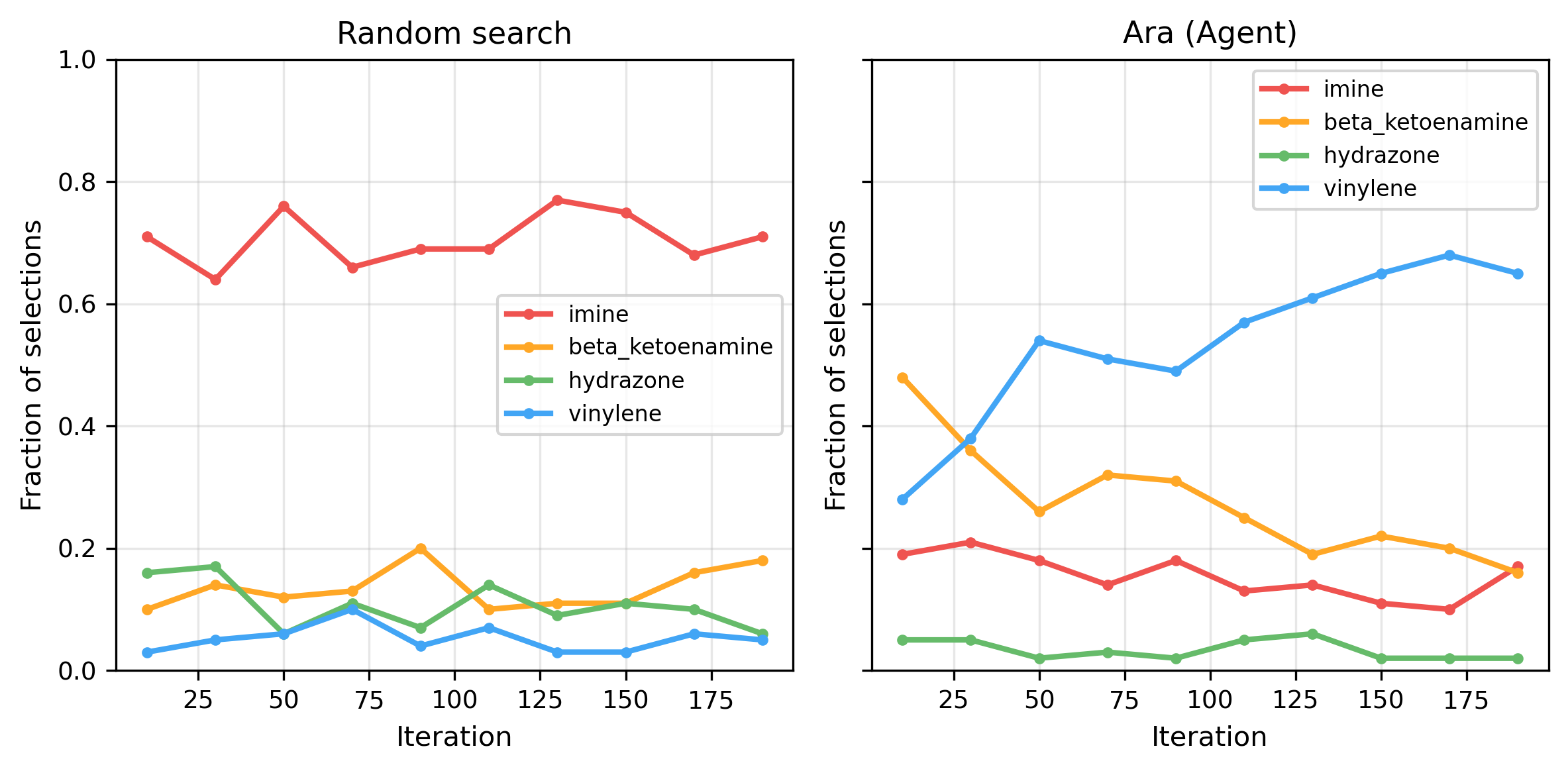}
\caption{Linkage-type distribution over search iterations. Fraction of candidate selections belonging to each linkage type (imine, $\beta$-ketoenamine, hydrazone, vinylene) in a rolling window of 20 iterations, averaged over five seeds. Left: random search maintains a roughly uniform distribution across linkage types throughout the run. Right: Ara (agent) shifts from a mixed distribution early in the search to a strong preference for vinylene and, to a lesser extent, $\beta$-ketoenamine by iteration 50, reflecting chemistry-aware prioritization of hydrolytically stable linkages (high S\textsubscript{CSI}) while exploring node and R-group choices to meet the band-gap and band-edge criteria.}
\label{fig:reasoning}
\end{figure}

Analysis of the linkage-type distribution over the full 200-iteration runs confirms the pattern quantitatively (Fig.~\ref{fig:reasoning}): the agent shifts from a mixed distribution early on to an overwhelming preference for vinylene
linkages (and, to a lesser extent, $\beta$-ketoenamine) by iteration~50, whereas random search maintains a uniform distribution throughout. The agent's three-level strategy, prioritise hydrolytically stable linkages, select electronically
appropriate nodes, then tune the band gap through R-group substitution, mirrors the hierarchical reasoning a human chemist would apply and is not naturally accessible to fingerprint-based BO without explicit categorical features for linkage chemistry.

\subsection*{Search space characterization and exploitation-exploration trade-off}

Exhaustive evaluation of all 820 candidates yielded 670 successful computations and identified 38 ground-truth hits, a global hit rate of 5.7\% (see Methods). The hits were concentrated in two linkage families: 30 vinylene-linked (79\%) and 8 $\beta$-ketoenamine-linked (21\%) COFs, with no imine- or hydrazone-linked candidates meeting all three criteria simultaneously (Supplementary Table~S3). The concentration of hits in these two families reflects the fundamental stability-activity trade-off: only vinylene and $\beta$-ketoenamine linkages achieve $S_\text{CSI} \geq 0.7$ while maintaining band gaps in the 1.8--2.2~eV window. Within the vinylene family, hit rates varied strongly by node: TFB and TFP-OH1 produced hits across nearly all R-groups, Tp was partially successful (9 of 10 R-groups, missing only SO$_3$H), and TFPT-ald yielded only a single hit (CF$_3$), as its strongly electron-withdrawing triazine core pushed the band gap below 1.8~eV for all other substitutions.

Using this ground truth, we computed oracle recovery, the fraction of the 38 unique hits discovered at least once, for each method (Supplementary Fig.~S2). BO achieved substantially higher oracle coverage than the agent, driven by the Gaussian process uncertainty term that actively pushes exploration into unsampled regions of fingerprint space. The agent's lower oracle recovery reflects its strong exploitation of a small number of productive vinylene-linked scaffolds: once it identified TFB $+$ DVB $+$ vinylene as a reliable hit source, it focused on R-group variants rather than exploring distant chemical families. Random search provided baseline diversity intermediate between the two. This exploitation-exploration trade-off has a practical
interpretation: in a budget-constrained experimental campaign where each evaluation corresponds to a DFT calculation or laboratory synthesis, the agent's
53\% per-iteration success rate is decisive; in a program seeking comprehensive characterization of the full design space, BO's systematic exploration is preferable. These are complementary, not competing, strengths.

\subsection*{Agent advantage is robust to scoring-function weights}

The composite stability index $S_\text{CSI}$ combines linkage stability ($S_\text{linkage}$), steric shielding ($S_\text{shielding}$), and hydrophobicity
($S_\text{hydrophobicity}$) with weights 0.50, 0.30, and 0.20, respectively. To assess whether the agent's advantage depends on this specific weight choice, we performed a sensitivity analysis across 30 weight triples in which the linkage weight varied from 0.3 to 0.7 and the remaining weight was distributed between shielding and hydrophobicity in increments of 0.1 (Supplementary Fig.~S1). For all 21 weight combinations where the search space contained at least 10 hits, the regime in which the comparison is statistically meaningful, the agent-to-random hit ratio ranged from 2.5$\times$ to 14.3$\times$, consistently exceeding the 2$\times$ threshold. The nine excluded triples had extreme shielding weights ($\geq 0.4$) that reduced the number of hits in the space to fewer than three, rendering ratio comparisons uninformative.

The robustness of the agent advantage across weight triples arises because the agent's primary learned strategy, preferring vinylene and $\beta$-ketoenamine linkages, produces high $S_\text{linkage}$ scores regardless of the exact weight assigned to this component. Stated differently, the agent's categorical reasoning about linkage chemistry is orthogonal to the specific numerical weighting within $S_\text{CSI}$, making the comparative result insensitive to
this methodological choice.

\section*{Discussion}

This work demonstrates that pretrained large language models encode chemical knowledge sufficient to guide efficient navigation of a multi-criteria COF design space. The Ara agent achieved a hit rate 11.5 times that of random search ($p < 0.01$), with the fastest convergence to the first hit among all tested methods (Table~\ref{tab:comparison}). Sensitivity analysis across 30 $S_\text{CSI}$ weight combinations confirmed that this advantage persists throughout the plausible weight space (Supplementary Fig.~S1), ruling out dependence on the specific scoring-function parameterization. These results establish that LLM chemical priors, specifically donor-acceptor theory, linkage stability knowledge, and structure-property reasoning, provide a viable foundation for materials discovery agents.

The agent's performance can be traced to specific chemical reasoning strategies that are visible in its output and verifiable against chemical intuition. First, the agent rapidly converged on non-hydrolysable vinylene (C$=$C) linkages, recognizing, without being told, that the absence of a hydrolyzable bond provides the highest achievable stability score. Second, the agent learned to avoid the strongly electron-withdrawing TFPT-ald node, correctly reasoning that the resulting donor--acceptor mismatch pushes the band gap below the 1.8~eV threshold; this mirrors the node dependence observed in the exhaustive evaluation, where TFPT-ald $+$ DVB $+$ vinylene produced only one hit out of ten R-groups. Third, the agent systematically explored R-group substitutions, OMe, NH$_2$, OH, Me, tBu, to fine-tune the band gap toward the 2.0~eV optimum. This three-level strategy (linkage $\to$ node $\to$ R-group) was discovered from evaluation feedback alone and mirrors the hierarchical reasoning a human chemist would apply. It is not naturally accessible to methods that encode molecules solely as numerical fingerprints without explicit categorical features for linkage chemistry. The agent was not, however, infallible: it occasionally proposed imine-linked candidates in early iterations despite their low $S_\text{linkage}$ scores (8 imine attempts in the first 50 iterations of seed~42, none yielding hits), and it revisited the TFPT-ald node a second time at iteration~13 despite its earlier failure (gap 1.75~eV, still below threshold). These episodes, together with the 23\% parse-failure rate, indicate that current LLMs remain imperfect chemical reasoners, though the overall strategy was sound.
The agent's performance can be traced to specific chemical reasoning strategies that are visible in its output and verifiable against chemical intuition. First, the agent rapidly converged on non-hydrolysable vinylene (C$=$C) linkages, recognizing, without being told, that the absence of a hydrolyzable bond provides the highest achievable stability score. Second, the agent learned to avoid the strongly electron-withdrawing TFPT-ald node, correctly reasoning that the resulting donor--acceptor mismatch pushes the band gap below the 1.8~eV threshold; this mirrors the node dependence observed in the exhaustive evaluation, where TFPT-ald $+$ DVB $+$ vinylene produced only one hit out of ten R-groups. Third, the agent systematically explored R-group substitutions, OMe, NH$_2$, OH, Me, tBu, to fine-tune the band gap toward the 2.0~eV optimum. This three-level strategy (linkage $\to$ node $\to$ R-group) was discovered from evaluation feedback alone and mirrors the hierarchical reasoning a human chemist would apply. It is not naturally accessible to methods that encode molecules solely as numerical fingerprints without explicit categorical features for linkage chemistry. The agent was not, however, infallible: it occasionally proposed imine-linked candidates in early iterations despite their low $S_\text{linkage}$ scores (8 imine attempts in the first 50 iterations of seed~42, none yielding hits), and it revisited the TFPT-ald node a second time at iteration~13 despite its earlier failure (gap 1.75~eV, still below threshold). These episodes, together with the 23\% parse-failure rate, indicate that current LLMs remain imperfect chemical reasoners, though the overall strategy was sound.

Exhaustive evaluation of all 670 successfully computed candidates identified 38
ground-truth hits (Supplementary Table~S3), enabling assessment of each method's
exploration breadth. The agent's high per-iteration hit rate reflects strong exploitation of a small number of productive chemical families, primarily
vinylene-linked COFs with TFB, TFP-OH1, and Tp nodes, rather than broad coverage of the full hit landscape. Bayesian optimization, driven by its Gaussian
process uncertainty term and built-in deduplication, achieved substantially higher unique-hit coverage (Supplementary Fig.~S2). These complementary behaviors suggest that budget-constrained experimental campaigns benefit from the agent's rapid convergence to high-quality candidates, while programs seeking comprehensive
characterization of the design space may prefer BO's systematic exploration. Hybrid strategies, using the agent for rapid initial convergence and BO for
subsequent diversification, represent a promising direction that could combine
the strengths of both approaches.

Approximately 23\% of agent iterations produced outputs that could not be parsed into valid candidate selections and were replaced with random draws. Rather than invalidating the comparison, this failure mode introduces a conservative bias: the agent effectively operated at approximately 77\% agency, with nearly one quarter of its budget consumed by random sampling. That it nonetheless achieved an 11.5-fold improvement over pure random search implies that the guided iterations were dramatically more efficient than chance. Improving LLM instruction-following and output formatting, through fine-tuning, constrained decoding, or structured-output APIs, would further enhance performance.

Several limitations should be acknowledged. First, the study uses a single LLM (Gemini-3-flash-preview); generalization across model families (e.g., GPT-5, Claude) and sizes remain to be established. Second, all evaluations are at the GFN1-xTB semiempirical level, while the transfer function provides a calibrated mapping to DFT band gaps (Spearman $\rho = 0.71$), periodic DFT confirmation of the top agent-discovered candidates is a necessary next step. Third, the search space is pre-enumerated (820 candidates); in open-ended generative design settings with $10^4$--$10^6$ candidates, the agent's categorical reasoning about linkage stability is expected to provide even greater advantage, though this remains to be demonstrated. Fourth, the agent operates with sampling-with-replacement, tending to re-evaluate known hits rather than explore; deduplication strategies that preserve chemical reasoning remain an open challenge. Finally, the composite stability index $S_\text{CSI}$ is a computational proxy that captures linkage chemistry, steric shielding, and hydrophobicity but does not account for the electronic influence of substituents proximal to the linkage bond on hydrolytic susceptibility, nor for framework crystallinity, both of which affect long-term durability under operating conditions. Experimental validation of hydrolytic stability under photocatalytic conditions is required to confirm the practical utility of the discovered candidates.

These findings open several avenues for future investigation. DFT validation of the top-ranked vinylene-linked candidates (TFP-OH1 $+$ DVB $+$ vinylene $+$ tBu, reward 0.921; TFB $+$ DVB $+$ vinylene $+$ OMe, reward 0.900) would establish whether the xTB-level predictions translate to accurate periodic band structures and band positions. Multi-model comparisons (Claude, GPT) would test whether the observed advantage reflects general LLM chemical priors or model-specific capabilities. Hybrid LLM--BO strategies could combine the agent's rapid convergence with BO's systematic exploration, potentially achieving both high sample efficiency and broad coverage. Multi-agent architectures, incorporating planner, critic, and memory components, may enable more sophisticated long-horizon reasoning. Expansion of the search space to include topology variation (honeycomb, kagome, square), pore-size engineering, and mixed-linker systems would test the agent's reasoning in higher-dimensional design spaces.
Finally, extension to other material classes, metal-organic frameworks, zeolites, and hybrid perovskites, would establish the generality of agent-guided
inverse design.

\section*{Methods}

\subsection*{COF search space definition}

The candidate space comprised 820 COF building-block combinations generated from a curated library of seven trigonal (C$_3$ symmetric) nodes, nineteen ditopic linkers, four linkage chemistries (imine, $\beta$-ketoenamine, hydrazone, vinylene), and ten aromatic R-group substituents (H, OH, NH$_2$, OMe, F, CF$_3$, NO$_2$, Me, tBu, SO$_3$H). Compatibility constraints reduced the combinatorial product. Imine linkages (C$=$N) form by Schiff base condensation between aldehyde and primary amine functional groups; in our library, these arise from aldehyde nodes paired with diamine linkers, or amine nodes paired with dialdehyde linkers. $\beta$-Ketoenamine linkages require 1,3,5-triformylphloroglucinol-type aldehydes bearing enolisable $\beta$-diketone motifs paired with primary amines; here, only the Tp node satisfies this requirement, paired with diamine linkers. Hydrazone linkages form between aldehyde and hydrazide functional groups; in our library, aldehyde nodes are paired with dihydrazide linkers. Vinylene linkages (C$=$C) form via Knoevenagel condensation between aldehydes and active methylene groups; in our library, aldehyde nodes are paired with DVB (1,4-bis(cyanomethyl)benzene).

The aldehyde nodes were TFB (1,3,5-triformylbenzene), TFPT-ald (triazine-2,4,6-tricarbaldehyde), TFP-OH1 (2-hydroxy-1,3,5-triformylbenzene), and Tp (1,3,5-triformylphloroglucinol); the
amine nodes were TAPB (1,3,5-tris(4-aminophenyl)benzene), TAPT
(2,4,6-tris(4-aminophenyl)-1,3,5-triazine), and TAPA
(tris(4-aminophenyl)amine). The breakdown by linkage type was: imine (aldehyde $+$ diamine) 440, imine (amine $+$ dialdehyde) 150, $\beta$-ketoenamine 110, hydrazone 80, and vinylene 40 candidates. Three node--linker combinations that consistently failed fragment assembly (TAPB $+$ Thieno-DA, TAPT $+$ OMe-PDA, TAPA $+$ Thieno-DA) were retained in the enumeration but excluded from analysis when assembly failed.

\subsection*{Fragment-based screening pipeline}

Each candidate was evaluated through a fragment-based computational pipeline.
First, a node--linker--node repeat-unit fragment was assembled from SMILES building
blocks using RDKit \cite{rdkit} reaction SMARTS specific to each linkage chemistry: imine
condensation (aldehyde $+$ amine $\to$ C$=$N), $\beta$-ketoenamine
tautomerization (via Tp $+$ amine), hydrazone formation (aldehyde $+$ hydrazide),
and Knoevenagel condensation (aldehyde $+$ vinyl $\to$ C$=$C). When the specified
R-group was not hydrogen, SMARTS-based substitution replaced one aromatic
hydrogen on the linker. The assembled fragment SMILES was converted to a
three-dimensional structure via RDKit's ETKDGv3 embedding with subsequent MMFF94
force-field minimization, yielding an initial geometry of typically 50--80 atoms.

Geometry optimization was performed with GFN1-xTB (version 6.7.1) \cite{grimme2017robust}
at the ``tight'' convergence level. The fundamental gap was computed via the
vertical ionization potential and electron affinity protocol
(\texttt{xtb -{}-vipea}), which performs delta-SCF calculations on the optimised
neutral geometry to obtain IP and EA values; the fundamental gap was taken as
IP $-$ EA. The linear transfer function (Eq.~\ref{eq:transfer}) was applied to
map xTB fundamental gaps onto the DFT scale, calibrated against literature DFT band gaps for 13 COFs spanning six linkage types (including boronate ester, boroxine, and triazine linkages not present in the search space; Supplementary Table~S2). The corrected gap was clipped to the range 0.5--6.0~eV to suppress unphysical outliers. The CBM position relative to NHE
was estimated as
\begin{equation}
  \text{CBM}_\text{vs NHE} = -\text{EA}_\text{xTB} + 2.62\;\text{V},
\end{equation}
where the offset was calibrated from three COFs with experimentally determined
band-edge positions: TpPa-1, TAPB-PDA, and sp2c-COF \cite{bi2019two,wang2018sulfone,wei2019semiconducting}.
The typical computational cost was 5--10 seconds per candidate for geometry
optimization and vipea on a single CPU core.

\subsection*{Composite stability index}

To encode hydrolytic stability as a continuous, differentiable score, we defined
a composite stability index $S_\text{CSI}$ combining three components:
\begin{equation}
  S_\text{CSI} = 0.50\,S_\text{linkage} + 0.30\,S_\text{shielding}
                + 0.20\,S_\text{hydrophobicity}.
  \label{eq:scsi}
\end{equation}

The linkage stability score $S_\text{linkage}$ was assigned based on the
well-established hydrolytic stability hierarchy of COF linkages
\cite{lyu2019porous,Kandambeth2012ConstructionCrystalline2DCOFs,Banerjee2013ChemicallyStableMultilayeredCONs,DeBlase2016MovingBeyondBoron,Zhuang2016TwoDimensionalConjugatedPolymerFramework}: imine 0.40, hydrazone 0.60,
$\beta$-ketoenamine 0.80, and vinylene 0.95. This ordering reflects the
underlying bond chemistry: imine C$=$N bonds undergo reversible hydrolysis;
hydrazone linkages are moderately protected by the adjacent amide;
$\beta$-ketoenamine bonds are stabilised by irreversible enol--keto
tautomerization and intramolecular C$=$O$\cdots$H--N hydrogen bonding
\cite{Kandambeth2012ConstructionCrystalline2DCOFs}; and vinylene C$=$C bonds are non-hydrolysable \cite{Zhuang2016TwoDimensionalConjugatedPolymerFramework}.

The steric shielding score $S_\text{shielding}$ quantified the local protection
of the linkage bond from nucleophilic attack by water. For each candidate, the
atoms forming the linkage bond were identified via SMARTS pattern matching on the
fragment SMILES: \texttt{[CH]=[N]} for imine, \texttt{[NH]-[CX3]=[OX1]} (carbonyl
C and O) for $\beta$-ketoenamine, \texttt{[CH]=[N][NH]} (C and first N) for
hydrazone, and \texttt{[CH]=[CH]} for vinylene. The midpoint of these atoms in
the xTB-optimised geometry served as the reference point, and $S_\text{shielding}$
was computed as the number of heavy atoms within a 4~\AA{} radius, normalised as
$\mathrm{clip}((\mathrm{count} - 5) / 15,\, 0,\, 1)$. When SMARTS matching failed
(e.g., due to unusual valence states), the molecular centre of mass was used as a
fallback.

The hydrophobicity score $S_\text{hydrophobicity}$ was computed as the Crippen
partition coefficient (SLogP) of the fragment SMILES using RDKit, normalised to
the range $[0, 1]$. Higher hydrophobicity disfavours water access to the linkage
bond, contributing to hydrolytic stability.

A sensitivity analysis across 30 weight triples ($S_\text{linkage}$ weight from
0.3 to 0.7; remainder distributed between $S_\text{shielding}$ and
$S_\text{hydrophobicity}$ in 0.1 increments) confirmed that the agent's advantage
over random search exceeded 2$\times$ for all weight combinations where the
search space contained at least 10 hits (Supplementary Fig.~S1).

\subsection*{Hit criteria and reward function}

A candidate was classified as a hit if it simultaneously satisfied three criteria:
(1) corrected band gap between 1.8 and 2.2~eV, targeting the visible-light
absorption window for photocatalytic water splitting; (2) CBM below 0~V versus
NHE, the thermodynamic requirement for proton reduction; and (3)
$S_\text{CSI} \geq 0.7$, ensuring adequate hydrolytic stability.

A continuous reward function was defined to guide the search methods:
\begin{equation}
  r = 0.35\,g_\text{score} + 0.30\,\text{cbm}_\text{score} + 0.35\,S_\text{CSI},
  \label{eq:reward}
\end{equation}
where
\begin{align}
  g_\text{score}         &= \exp\!\left(-\tfrac{1}{2}
    \left(\frac{g - 2.0}{0.3}\right)^2\right),\\
  \text{cbm}_\text{score} &= \frac{1}{1 + \exp\!\left(5.0\times\text{CBM}_\text{vs NHE}\right)}.
\end{align}
The reward ranged from 0 to approximately 1, with values above 0.85 typically
corresponding to hits.

\subsection*{Search methods}

Three search methods were compared, each evaluated over 200 iterations with five
random seeds (42, 123, 456, 789, 2024) using a shared evaluation cache to ensure
identical computational results for identical candidates.

\paragraph{Random search.}
Candidates were sampled uniformly at random (with replacement) from the
820-member candidate pool. This served as the null-hypothesis baseline.

\paragraph{Bayesian optimization (BO).}
A Gaussian process (GP) surrogate was fitted on 2048-bit Morgan fingerprints
(radius 2) of assembled fragment SMILES, computed via RDKit. Features were
standardized with a zero-mean, unit-variance scaler. The GP used a Mat\'{e}rn
kernel ($\nu = 2.5$, length scale 1.0, bounds $[10^{-3}, 100]$) with normalized
targets, fitted with five random restarts of the L-BFGS-B optimizer \cite{Byrd1995LBFGSB}. The first
20 iterations used random initialization; subsequent iterations selected the
candidate with the highest expected improvement (EI) among unevaluated
candidates. Evaluated candidates were tracked and excluded from subsequent
acquisition-function maximization, ensuring that each iteration evaluated a novel
candidate. This built-in deduplication is a standard feature of BO that
contributes to its exploration efficiency. By contrast, the LLM agent operated
with sampling-with-replacement, preserving its natural tendency to exploit
high-reward chemical families; the trade-off between this exploitative behavior
and exploration breadth is examined in the Discussion.

\paragraph{LLM agent (Ara).}
The agent used Google's Gemini-3-flash-preview model \cite{team2023gemini} in JSON output
mode (temperature 0.3, max 1024 output tokens). At each iteration, the agent
received a prompt containing: (a) a system instruction describing the COF design
task, available building blocks, compatibility constraints, and chemical principles
(donor--acceptor theory, conjugation effects, stability--activity trade-off); (b)
the results of the last 10 evaluations and the top 5 by reward from the full
history; and (c) a request to select one (node, linker, linkage, R-group)
combination with 1--3 sentences of chemical reasoning. The agent's response was
parsed for the four building-block identifiers; when parsing failed or the response was empty, a random candidate was substituted (contributing to the
reported parse-failure rate). No explicit deduplication was enforced: the agent
was free to re-evaluate previously selected candidates, relying on the soft
guidance in the prompt (``choose a NEW combination you have not tried before if
possible'') and the list of previously evaluated candidates to encourage diversity.
The consequences of this sampling-with-replacement design for exploration diversity
are examined through oracle recovery analysis in the Discussion.

\paragraph{Statistical testing.}
Pairwise comparisons of cumulative hits across seeds were performed using the
one-sided Mann--Whitney $U$ test, with significance declared at $p < 0.05$.

\subsection*{Ground-truth evaluation}

To establish a ground-truth benchmark, all 820 candidates in the search space
were submitted to the fragment-based screening pipeline. Of these, 670 (81.7\%)
were successfully evaluated; the remaining 150 failed at fragment assembly or
xTB convergence. Among the 670 evaluated candidates, 38 satisfied all three hit
criteria (Supplementary Table~S3), yielding a global hit rate of 5.7\%. Oracle
recovery was defined as the fraction of these 38 ground-truth hits discovered at
least once (unique hits only, with no double-counting of repeated evaluations)
within a method's 200-iteration budget.

\paragraph{Reasoning trace collection.}
To obtain interpretable reasoning examples (Table~\ref{tab:traces}), a separate
20-iteration run was performed using the same model (Gemini-3-flash-preview) and
system prompt, with the maximum output-token limit increased from 1024 to 4096 to
avoid truncation of reasoning text and the temperature raised to 0.5 to encourage
more diverse reasoning. All other generation parameters (JSON output mode, safety
settings) were unchanged. This run was used for qualitative analysis only and did
not contribute to the quantitative comparison.

\section*{Data availability}
Data supporting the findings of this study are available in the following locations:

\begin{itemize}
    \item \textbf{Building blocks library} (nodes, linkers, linkages, R-groups): included in this repository as `data/building\_blocks\_library.json`.
    \item \textbf{Calibration dataset} (xTB vs DFT band gaps for 13 COFs): `data/calibration\_dataset.csv`.
    \item \textbf{CBM calibration} (conduction-band minimum vs NHE): `data/cbm\_calibration.json`.
    \item \textbf{Comparative study outputs} (run results, oracle summary, full evaluation)
\end{itemize}

All other data generated or analysed during this study are either included in this repository or can be reproduced using the provided code and the above data files.

\section*{Code availability}
The code that supports the findings of this study is available at https://github.com/Iman-Peivaste/Ara . The repository includes the fragment-based screening pipeline (main entry point `main.py`), search methods (random, Bayesian optimisation, and LLM agent), calibration and analysis scripts, and the building blocks library and calibration datasets required to reproduce the results. The pipeline relies on the GFN1-xTB implementation in the xtb programme (see Methods), RDKit for molecular handling, and (for COF construction) the Supramolecular ToolKit (stk). Instructions for environment setup and reproduction of the comparative study and figures are provided in the repository README.

\section*{Author Contributions}

Iman Peivaste: Conceptualization, developing the initial concept and workflow, framework design, writing the original draft, visualization, methodology, investigation, and formal analysis. 

Nicolas Boscher: Writing, review, editing, resources, methodology, and formal analysis.

Ahmed Makradi: Writing, review, editing, methodology, and formal analysis.

Salim Belouettar: Supervision, funding acquisition, conceptualization (research direction and framing), writing – review \& editing, methodology, resources.

\section*{Competing Interests}

The authors declare that they have no known competing financial interests or personal relationships that could have appeared to influence the work reported in this paper.

\section*{Acknowledgements}

This work was funded by the Luxembourg Fonds National de la Recherche (FNR) through the grant PRIDE21/16758661/HYMAT.  


\clearpage
\bibliographystyle{naturemag}. 
\bibliography{refs}

\end{document}